\documentclass[aps,twocolumn,amsmath,amssymb,floatfix,prb]{revtex4}
\usepackage{graphicx}
\usepackage{dcolumn}
\usepackage{bm}
\usepackage{amsfonts}
\usepackage{graphicx}
\usepackage{dcolumn}
\usepackage{bm}
\usepackage{amsmath}
\usepackage{amssymb}

\begin{document}

\title{Ferromagnetism in chiral multilayer 2D semimetals} 
\author{Hongki Min$^{1,2}$, E. H. Hwang$^{1,3}$, and S. Das Sarma$^1$}
\affiliation{$^1$Condensed Matter Theory Center and Joint Quantum Institute, 
Department of Physics, University of Maryland, College Park,
Maryland  20742-4111 \\
$^2$Department of Physics and Astronomy, Seoul National University,
Seoul 08826, Korea \\
$^3$SKKU Advanced Institute of Nanotechnology and Department of
Physics, Sungkyunkwan  University, Suwon, 16419, Korea}


\begin{abstract}
We calculate the temperature dependent long-range magnetic coupling in the presence of dilute concentrations of random magnetic impurities in chiral multilayer two-dimensional semimetals, i.e., undoped intrinsic multilayer graphene. Assuming a carrier-mediated indirect RKKY exchange interaction among the well-separated magnetic impurities with the itinerant carriers mediating the magnetic interaction between the impurities, we investigate the magnetic properties of intrinsic multilayer graphene using an effective chiral Hamiltonian model. We find that due to the enhanced density of states in the rhombohedral stacking sequence of graphene layers, the magnetic ordering of multilayer graphene is ferromagnetic in the continuum limit. The ferromagnetic transition temperature is calculated using a finite-temperature self-consistent field approximation and found to be within the experimentally accessible range for reasonable values of the impurity-carrier coupling.
\end{abstract}

\maketitle


\section{Introduction}

Multilayer graphene with an additional layer degree of freedom 
(in addition to the spin and pseudospin intralayer index)
has recently attracted a great deal of attention for its fundamental properties as well as for its potential applications.\cite{dassarma_rmp,Raza,yacoby,multilayer, min2012,min2008,trilayer} Multilayer graphene is not a simple extension of monolayer 
graphene (since it has its own characteristic layer-number-dependent band structure and symmetry properties),
and could open the possibility of engineering electronic properties by tuning the stacking arrangement. One important salient feature of multilayer graphene is the enhancement of the electronic density of states (DOS) as the number of graphene layers increases. 
As a consequence, the electronic screening becomes more important with increasing layers.\cite{min2012} Since the energy band structure of multilayer graphene is very sensitive to its stacking sequence, the screening properties depend strongly on the stacking arrangements in multilayer graphene.\cite{min2012,min2008}
Each type of rhombohedral multilayer graphene (i.e., $J$-graphene) with the layer number index $J=1$, 2, 3, 4, 5... is a distinct 2D material tuned by $J$ except that all of them are 2D gapless semimetals with the chemical potential precisely pinned at the touching point of conduction and valence bands.  The most well-known $J$-grahenes are monolayer graphene (MLG) with $J=1$ and bilayer graphene (BLG) with $J=2$, but trilayer ($J=3$) and even higher-layer ($J>3$) graphenes have also been studied in the laboratory.\cite{trilayer} In fact, the $J$ going to infinity limit (i.e., infinite-layer graphene) is graphite.

The current theoretical work is on the $J$-dependent magnetic properties of intrinsic (i.e., undoped) multilayer graphene with no free carriers in the conduction or valence band at $T=0$.
We study finite temperature response (screening) functions of multilayer graphene, and their consequences for the $J$-dependent magnetic properties induced by magnetic impurities
 through the 
Ruderman-Kittel-Kasuya-Yosida (RKKY) interaction.\cite{rkky_c,rkky_cc} In the presence of a dilute concentration of magnetic impurities in nonmagnetic metals the effective exchange interaction between the impurities is induced as the second-order perturbation with respect to the direct exchange interaction between the magnetic impurity and the itinerant electrons of the host (i.e., the magnetic impurities experience a long-range indirect exchange interaction mediated by conduction electrons, known as the RKKY interaction). 
Such an indirect nonlocal carrier-mediated  RKKY interaction between two impurities could be ferromagnetic or antiferromagnetic depending on their spatial separation because the interaction is oscillatory due to the sharpness of the Fermi surface.  This indirect RKKY interaction exists in addition to any possible direct exchange interaction among the magnetic impurities which may arise due to their direct wave function overlap.
Because the RKKY interaction is mediated by host electrons (or holes), the enhanced DOS of multilayer graphene can lead to an increase of the effective inter-impurity magnetic coupling with increasing $J$, which may induce robust magnetic ordering
for larger values of the layer number. The goal of the current paper is to theoretically predict such magnetic ordering in multilayer graphene as a function of layer number.

In general, intrinsic graphene does not have any permanent magnetic moments in the bulk, but local magnetic moments can be introduced by extrinsic doping. 
Doping by suitable magnetic impurities could introduce such local moments, but these local moments cannot order spontaneously unless there is an inter-impurity magnetic interaction. If the impurities are far apart, i.e., the doping is dilute (which is the only situation considered in the current work), then the direct exchange interaction among the impurities is basically zero since their wave function overlap is exponentially small.
It is also now well-accepted that graphene (or generally, multilayer graphene) is not intrinsically (i.e., in the absence of any doping) magnetic, i.e., there is no spontaneous graphene magnetic ordering of any kind unless local magnetic moments are extrinsically introduced by magnetic impurities, vacancies, or edges.\cite{rkky00}  The important question addressed in this paper is whether a dilute (i.e., well-separated impurities) concentration of magnetic dopants could induce magnetism in graphene through the RKKY mechanism with direct exchange playing no role whatsoever.  We emphasize that the dilute limit is very different from the dense (or Kondo lattice) limit where the magnetic impurities themselves form a lattice (or are in almost every unit cell of the host lattice) since direct interaction among the impurities as well as any modification of the graphene band structure induced by the impurities can be neglected. The only interaction to be considered in this dilute limit is the indirect RKKY interaction, and a continuum approximation should suffice. The dilute approximation with well-separated dopants makes the graphene situation here very similar to the extensively studied \cite{DMS_ssc,DMS_rmp} 
diluted magnetic semiconductor (DMS) materials where also the primary mechanism driving ferromagnetism in the semiconductor is thought to be the indirect RKKY interaction.  This is the physics we investigate in the current work. The main qualitative difference between diluted-magnetic-graphene (DMG) we consider here and the well-studied DMS is that we must deal with an undoped gapless semimetal (and not a doped semiconductor) and we must incorporate the layer number for multilayer graphene.

It is known that the effective RKKY interaction in graphene induced by the exchange coupling between local magnetic moments and conduction electrons (or holes) behaves differently from the ordinary two dimensional (2D) systems because of the chiral gapless nature of graphene.\cite{brey2007,wunsch} 
In particular, the strength of RKKY coupling in monolayer graphene decays faster spatially than in ordinary 2D electron systems due to chirality (i.e., the suppression of the 2$k_F$ scattering, where $k_F$ is the Fermi wave vector) in graphene.\cite{brey2007,wunsch}

One may wonder if RKKY interaction, with its long-range spatial Friedel oscillations, is capable of inducing magnetic ordering since the inter-impurity interaction may be of random sign (ferromagnetic or antiferromagnetic) depending on their spatial locations. This is certainly the case in ordinary metals or semiconductors (either 2D or 3D)\cite{rkky_cc} in the presence of a high concentration of magnetic impurities where the effective inter-impurity interaction will be randomly ferromagnetic and antiferromagnetic, leading to considerable frustration in the Hamiltonian (and perhaps therefore a glassy ground state with no obvious long-range order).  But for a low or dilute impurity concentration (as in DMS) where the impurities are on the average far from each other, the effective inter-impurity RKKY interaction is on the average mostly ferromagnetic, and then the possibility of  magnetic ordering arises, albeit with perhaps a low transition temperature (as in DMS) because of the generally weakened average inter-impurity RKKY interaction.\cite{DMS_ssc,DMS_rmp}  In multilayer graphene, layer index and chirality introduce novel RKKY physics necessitating a specific analysis to search for possible magnetic ordering of DMG.  In particular, multilayer graphene RKKY interaction does not change sign (in spite of Friedel oscillations) for $J \ge 3$ indicating a strong tendency toward a long-range ordering of the magnetic impurities induced by RKKY coupling.  Another fundamental difference between intrinsic graphene and ordinary metals or doped semiconductors is the fact that undoped graphene is a gapless semiconductor or a semimetal with no free carriers at $T=0$ since the chemical potential separates a filled valence band touching an empty conduction band.  Taken together, all these differences between chiral intrinsic DMG and regular DMS  imply that our intuition based on the substantial body of DMS literature  is a poor guide to understanding how graphene magnetism may arise from RKKY physics.  In this paper, we provide a complete picture based on a continuum mean field theory, which should be qualitatively and semiquantitaively valid in the dilute impurity density limit.

Magnetic properties of graphene have been studied, and in particular, there have been several studies of RKKY interaction in graphene focusing on the possibility of magnetic ordering
of dopant magnetic impurities at zero temperature.\cite{brey2007,wunsch, rkky1,rkky2,rkky3,rkky4,rkky5,rkky6, kogan2011, jiang2012} 
However, a systematic study of RKKY interaction in multilayer graphene as a function of layer index has not been undertaken, and, in addition, finite temperature, disorder and finite carrier mean-free path effects on RKKY interaction have not yet been studied theoretically.
In this paper, we calculate the magnetic properties of multilayer graphene
with the magnetic impurities located at the interface between graphene and substrate without breaking any symmetry in the graphene layer. We calculate the temperature dependence of the RKKY interaction in multilayer graphene  in order to develop a self-consistent field theory to study long-range finite-temperature magnetic ordering. We show that the enhanced DOS in rhombohedral stacking allows ferromagnetic ordering of the magnetic impurities at experimentally accessible temperatures, particularly for higher values of $J$. Our results indicate that
the magnetic impurity induced ferromagnetic ordering is possible in semimetallic multilayer graphene 
arising from the RKKY indirect interaction in the dilute impurity limit.  Ferromagnetism in DMG as predicted in our theory, particularly for larger layer numbers, could in principle lead to graphene spintronics if our predictions are validated experimentally.

This paper is organized as follows. In Sec. II, we describe our model and theoretical approaches based on chiral 2D electron systems. In Sec. III,
we give the calculated results for RKKY interaction and the ferromagnetic transition temperatures of DMG. We conclude in Sec. IV with a discussion of the momentum cutoff effects on the effective coupling.

\section{Model}

To  study indirect magnetic interaction between quenched local moments in multilayer graphene, 
we consider the indirect exchange interaction between magnetic impurities to be of the RKKY form, i.e., carrier-mediated effective magnetic interaction. Indirect exchange interaction between magnetic moments is determined by the electronic structure of the relevant system. To describe electron states in multilayer graphene, we consider the Hamiltonian of noninteracting electrons in the form of a two-band pseudospin Hamiltonian for 2D chiral quasiparticles. Thus, multilayer graphene near the band-touching Dirac point can be described by a set of chiral 2D electron systems (C2DESs) and the Hamiltonian with the chirality index $J$ (which also represents the number of layers in rhombohedral multilayer graphene) is of the form \cite{min2012, min2008}
\begin{equation}
H_J(\bm{k})=t_{\perp}\left(
\begin{array}{cc}
0 & \left({\hbar v_0 k_{-}\over t_{\perp}}\right)^J \\
\left({\hbar v_0 k_{+}\over t_{\perp}}\right)^J & 0 \\
\end{array}
\right),
\end{equation}
where $k_{\pm}=k_x\pm i k_y$, $v_0$ is the effective in-plane Fermi velocity, and $t_{\perp}$ is the nearest-neighbor interlayer hopping.
The Hamiltonian has an energy spectrum given by $\varepsilon_{\lambda,\bm{k}}=\lambda t_{\perp} \left(\hbar v_0 |\bm{k}| \over t_{\perp}\right)^J$ and the corresponding eigenfunctions are $\left|\lambda,\bm{k}\right>={1 \over \sqrt{2}}\left(\lambda,e^{i
  J\phi_{\bm{k}}}\right)$, where $\phi_{\bm{k}}=\tan^{-1}(k_y/k_x)$
and $\lambda=\pm 1$ for conduction (valence) band energy states, respectively. 
We consider intrinsic multilayer semimetallic graphene with the Fermi energy precisely at the Dirac point which we take to be the zero of energy.

The carrier mediated RKKY indirect exchange interaction describing the effective magnetic interaction between local moments induced by the free carrier spin polarization is proportional to the static carrier susceptibility. The finite temperature static susceptibility can be calculated using the finite temperature Fermi distribution function as \cite{dassarma_rmp,hwang_rpa}
\begin{equation}
\chi(\bm{q},T)=-g\sum_{\lambda,\lambda'}\int {d^2 k \over (2\pi)^2} {f_{\lambda,\bm{k}}-f_{\lambda',\bm{k}'} \over \varepsilon_{\lambda,\bm{k}}-\varepsilon_{\lambda',\bm{k}'}} F_{\lambda,\lambda'}(\bm{k},\bm{k}'),
\label{eq:bare_polarization}
\end{equation}
where $g=g_{\rm s} g_{\rm v}$ is the total degeneracy factor ($g_{\rm s}=g_{\rm v}=2$ are spin and valley degeneracy factors, respectively), 
$f_{\lambda,\bm{k}}=1/[\exp(\varepsilon_{\lambda,\bm{k}}/k_BT) + 1]$ is the finite temperature Fermi distribution function for the band index $\lambda=\pm1$ and wave vector $\bm{k}$, 
$F_{\lambda,\lambda'}(\bm{k},\bm{k}')$ is the square of the wave-function overlap between $\left|\lambda,\bm{k}\right>$ and
$\left|\lambda',\bm{k}'\right>$ states, and $\bm{k}'=\bm{k+q}$.
For the chiral electron with the chirality index $J$, $F_{\lambda,\lambda'}(\bm{k},\bm{k}')={1\over 2}\left[1+\lambda\lambda'\cos J(\phi_{\bm k}-\phi_{\bm k}')\right]$.

For the undoped intrinsic case, in which the chemical potential is located at the Dirac point for all temperatures, Eq.~(\ref{eq:bare_polarization}) can be expressed as 
\begin{eqnarray}
\chi(q,T) & = & D_J(q) [I_J^{+}(q,T) + I_J^{-}(q,T)] \nonumber \\
              & = & \chi^+_J(q,T) + \chi^-_J(q,T),
\end{eqnarray}
where $D_J(q)$ corresponds to the DOS with a wave number $q$ which is given by
\begin{equation}
\label{eq:intrinsic_bare_polarization}
D_J(q)  = {g q^{2-J} \over 2\pi J t_{\perp}(\hbar v_0/t_{\perp})^J}, 
\end{equation}
and
\begin{eqnarray}
I_J^{\pm}(q,T) &= &J\int_0^{\infty} x dx  \int_0^{2\pi} \frac{d\phi}{2\pi} \frac{1 \pm \cos (J\theta)}{x^J \pm (x')^{J}} \nonumber \\
&\times& \left [ \tanh \frac{T_J(qax)^J}{T}  \pm \tanh \frac{T_J(qax')^J}{T}  \right ],
\end{eqnarray}
where $x'=\sqrt{1+2x\cos\phi+x^2}$, $\cos\theta = (x+\cos \phi)/x'$, $T_J = ({t_{\perp}}/{k_B})({\hbar v_0}/{t_{\perp}a})^J$, and $a$ is the lattice constant of graphene.
At zero temperature ($T=0$), $I_J^-$, which corresponds to the intraband transition of electrons, vanishes for all $q$  (since the conduction band is completely empty and the valence band completely full) and only interband transition, $I_J^+$, contributes to the susceptibility,  
and we have $\chi(q,T=0) = \chi^+_J(q)= D_J(q) I_J^+(q)$ \cite{min2012}.
As $q \rightarrow 0$,  we have $\chi^+_J(q,T=0) \propto q^{2-J}$ and  therefore, for $J\ge 3$, the static susceptibility diverges at long wavelength.

At finite temperatures ($T>0$), it is interesting to notice that the interband susceptibility behaves like $\chi^+_J(q,T) \propto q^2$ as $q\rightarrow 0$ for all $J$. Thus, the $q=0$ singular behavior of the interband susceptibility at $T=0$ for $J\ge 3$, i.e., $\chi_J^+(q\rightarrow 0,T=0) \rightarrow \infty$, disappears at finite temperatures. In addition, intraband transitions 
contribute to the susceptibility at finite temperatures due to thermal particle-hole excitations in the semimetal, and thus, at small $q$ the total susceptibility comes entirely from the intraband transition at finite $T$, $\chi_J^-$,   due to the vanishing of long wavelength interband contribution. Especially, we find that $\chi_J^-(q=0,T) \propto T^{2-J}$ for $J \le 2$ and $\chi_J^-(q=0,T) \propto 1/T$ for $J\ge 3$. Thus, the total susceptibility at $q=0$, $\chi(0,T)$, increases with temperature only for $J=1$. For $J=2$, $\chi(0,T)$ is constant for all temperatures, and it decreases with increasing temperature for $J \ge 3$. We will discuss the implications of these temperature dependences 
for $J$-dependent long-range magnetic ordering.

\begin{figure}[t]
\includegraphics[width=1.0\linewidth]{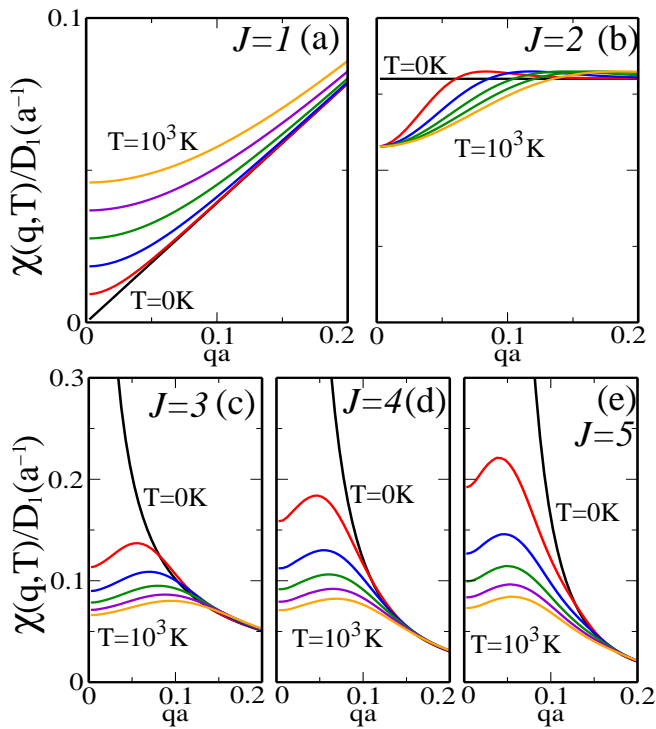}
\caption{The calculated finite temperature static polarizability $\chi(q,T)$ as a function of wave vector for various temperatures $T=0$, 200, 400, 600, 800, and 1000 K, and for different chiralities (a) $J=1$, (b) $J=2$, (c) $J=3$, (d) $J=4$, and (e) $J=5$. 
Here $D_1(a^{-1})=g/(2\pi t_{\perp}a^2)(t_{\perp}a/\hbar v_0)$ is given in Eq.~(\ref{eq:intrinsic_bare_polarization}) with $J=1$ and $q=q^{-1}$, and $a$ is the length scale of the system (i.e., lattice constant) and $a=2.46$\AA \ is used. 
} 
\label{fig:polarization}
\end{figure}

It is not possible to obtain the susceptibility function analytically for all $q$ at finite temperatures. Thus, we calculate the finite temperature static susceptibility numerically. Figure \ref{fig:polarization} shows the calculated static susceptibility $\chi(q,T)$ as a function of wave vector for several temperatures. For comparison, we normalize the susceptibility for different $J$ by the $J=1$ DOS at $q=1/a$, $D_{J=1}(q=a^{-1})$, where $a$ is the lattice constant of graphene. As shown in Fig.~1(a) the $J=1$ susceptibility increases linearly with temperature at small $q$. For $J=2$ we find that $\chi(0,T=0) \neq \chi(0, T\rightarrow 0)$ and $\chi(0,T=0)/\chi(0,T\neq 0)=\ln 4$ for all finite temperatures. The finite temperature susceptibility at $q=0$ is independent of the temperature
for $J=2$ as discussed above. 
It is interesting to compare this behavior with the asymptotic form for the corresponding nonchiral regular 2D electron gas susceptibility which exponentially decreases from its zero temperature value, $\chi(q=0,T) \approx \chi(q=0,T=0)  [ 1- e^{-T_F/T}]$, where $T_F$ is the Fermi temperature of the system.\cite{dassarma_rmp,hwang2009} For $J \ge 3$, even though the zero temperature susceptibility is infinity at $q=0$, the finite temperature susceptibility is finite and $\chi(0,T)$ decreases inverse linearly with temperature, $1/T$. Overall 
$\chi(q,T)$ for $J\ge 3$ manifests a similar behavior.

\section{RKKY interaction and effective magnetic coupling}

The interaction between a localized spin ${\bm S}_i$ located at ${\bm r}_i$ and an itinerant electron spin ${\bm s}$ at ${\bm r}$, i.e.,  $V({r})=J_{\rm ex}{\bm S}_i\cdot {\bm s} \ \delta({\bm r}_i-{\bm r})$ with an exchange coupling $J_{\rm ex}$,  accounts for the interaction between magnetic impurities. 
In general, $J_{\rm ex}$ is an unknown parameter in our theory which must be determined experimentally or from a separate first principles calculation beyond the scope of the current work.
Then the effective Hamiltonian describing magnetic interaction between two classical Heisenberg spins ${\bm S}_i$ and ${\bm S}_j$ located at ${\bm r}_i$ and ${\bm r}_j$, respectively, is given by \cite{priour,kittel_book}
\begin{equation}
H= -\sum_{i,j}J_{\rm RKKY} ({\bm r}_i-{\bm r}_j) {\bm S}_i\cdot {\bm S}_j,
\end{equation}
where 
\begin{equation}
J_{\rm RKKY} ({\bm r},T) = {\left[J_{\rm ex} a^2 \right]^2 \over 4} \chi({\bm r},T).
\end{equation} 
Note that the `classical moment' approximation here is justified by the large moments of the magnetic impurities typically used for magnetic doping and quantum fluctuations in the magnetic impurity are neglected as being small.  We ignore all complications associated with Kondo physics and other quantum strong correlation aspects assuming that the long-range Heisenberg model is the appropriate model for describing magnetic ordering for DMG.
We essentially assume that the relevant Kondo temperature is much less than the RKKY temperature scale in the system.\cite{Kondo}
The RKKY interaction is related to the range function which is defined by the Fourier transform of the static susceptibility, i.e., $\chi({\bm r},T)=\sum_{\bm q} \chi({\bm q},T)$ and in 2D it is given by
\begin{equation}
\chi({\bm r},T)=\int_0^{\infty} {q dq \over 2\pi} J_0(qr)\chi({\bm q},T),
\label{transform}
\end{equation}
where $J_0(x)$ is the Bessel functions of the first kind. Note that even though the Fourier transform of Eq.~(\ref{transform}) is well defined for $J=2$ and 3, it requires an ultraviolet cutoff for $J=1$ and an infrared cutoff for $J \geq 4$. These large and small momenta regularizations are necessary for obtaining meaningful nonsingular results.
In the following results, we set the infrared momentum cutoff as $q_c^{(l)}=0.01/a$ and the ultraviolet momentum cutoff as $q_c^{(h)}=1/a$, where $a$ is the typical length scale of the system, i.e., a lattice constant of graphene, and we use $a=0.246$ nm in our numerical calculations. 
While the large momentum ultraviolet cutoff (inverse lattice constant) for $J=1$ is natural in a continuum theory, the infrared low momentum cutoff is not usual in solid state physics.
We provide the details on these regularizations and their possible effects in the discussion section (see Sec. IV).

\begin{figure}
\includegraphics[width=1.0\linewidth]{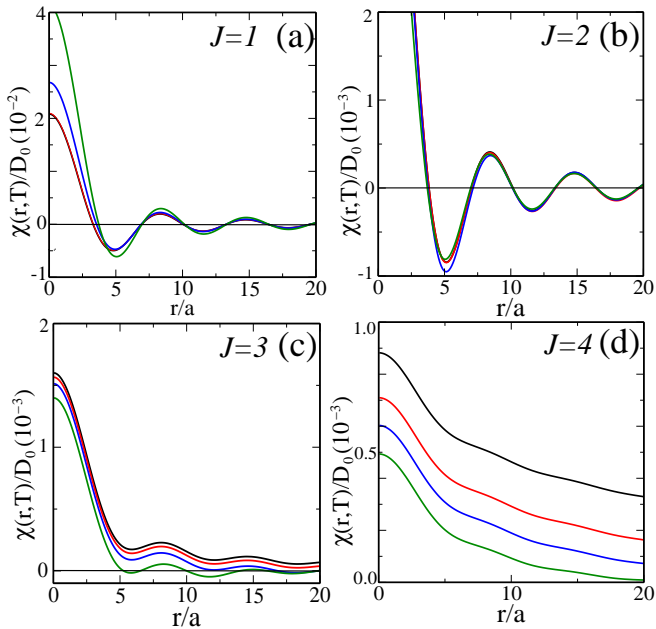}
\caption{The RKKY range function $\chi({\bm r})$ as a function of distance 
for different chiralities (a) $J=1$, (b) $J=2$, (c) $J=3$, (d) $J=4$. In each figure the different curves represent different temperatures $T=0$, 10, 100, 1000 K [from bottom to top in (a), and from top to bottom in (b), (c), (d)].
Here $D_0 = D_1(a^{-1})/a^2$.
In this calculation the infrared momentum cutoff $q_c^{(l)}=0.01/a$ for $J =4$ and the ultraviolet momentum cutoff $q_c^{(h)}=1/a$ for $J=1$ are used.} 
\label{fig:range_function}
\end{figure}

Figure \ref{fig:range_function} shows the RKKY range functions $\chi({\bm r},T)$ for $J=1,2,3,4$ and for different temperatures. For $J=1$ the range function oscillates for all temperatures, and its magnitude increases with temperature. For $J=2$ the range function is almost independent of temperature. For $J\ge 3$ the magnitude of the range function decreases with temperature. These behaviors for different $J$ can be understood from the temperature dependence of the susceptibility shown in Fig.~1. More importantly, we find that for $J=1,2$ the range functions alternate between positive and negative values while, for $J\geq 3$, it always remains positive at $T=0$. Due to the suppression of large $q$ contribution for higher values of layer number $J$, 
the range functions for $J\ge 3$ do not have oscillations. Hence, there are no competing ferro- and antiferromagnetic couplings for $J\ge 3$, and the magnetic impurity moments are expected to be ferromagnetically aligned since there is no frustration in the RKKY coupling.

The carrier-mediated RKKY interaction induced indirect exchange interaction [Eqs.~(6) and (7)] describes the effective magnetic interaction between local magnetic moments induced by the free carrier spin polarization.
The effective temperature-dependent coupling is then given by the spatial average of the $J_{\rm RKKY}$: 
\begin{equation}
J_{\rm eff}(T)  ={1\over \Omega_{\rm unit}} \int d^2 r J_{\rm RKKY} ({\bm r},T),
\end{equation}
where $\Omega_{\rm unit}$ is the area of a unit cell. The  effective temperature dependent coupling can be expressed in the dimensionless form
\begin{equation}
{J_{\rm eff}(T) \over J_{\rm eff}^{(0)}} = { 1 \over D_1(a^{-1})} \int r dr \chi({\bm r},T) 
\label{jeff}
\end{equation}
where $J_{\rm eff}^{(0)}={\left[J_{\rm ex} a^2 \right]^2} \times {2\pi D_1(a^{-1})/ 4 \Omega_{\rm unit}}$ is a magnetic coupling constant, which is proportional to the square of the local exchange coupling, but
independent of the chirality index $J$ and temperature $T$.  
All the interesting physics of chirality index $J$ and temperature $T$ enters through the integral in Eq.~(\ref{jeff}) which depends nontrivially and nonlinearly on both $J$ and $T$. We also include disorder effects phenomenologically through a finite carrier mean-free path by including an exponential cutoff in the range of the RKKY interaction, which allows us to take into account the dependence of the magnetic behavior of multilayer graphene on the carrier transport properties. 
In the presence of (nonmagnetic) impurity  scattering 
the RKKY interaction range is cut off at long distances, and we include this physics through
an exponential spatial damping at distances larger than a characteristic length scale of the order of the carrier transport mean-free path\cite{priour,eggenkamp1995,disorder,disorder1986}. Thus the effective coupling can be modified as
\begin{equation}
J_{\rm eff}  =
\begin{cases}
{1\over \Omega_{\rm unit}} \int d^2 r J_{\rm RKKY} ({\bm r}),  & \text{$(r < R)$} \\
{1\over \Omega_{\rm unit}} \int d^2 r J_{\rm RKKY} ({\bm r}) e^{-{{r-R}\over R}}. & \text{$(r > R)$}
\end{cases}
\end{equation}
Here the exponential cutoff $R$ is introduced to take into account the finite mean-free path due to scattering by nonmagnetic disorder in the semimetal. In the calculation, we use $R=100 a=24.6$ nm, which is a characteristic scale of the mean-free path, and the choice of $R$ ($<300a$) does not change our results qualitatively. 
We note that the appropriate mean free path ($R$) here is the one corresponding to undoped intrinsic multilayer graphene near the Dirac point, which depends on $J$ and $T$ (and should be taken from transport data).  We use the length cutoff parameter $R$ just as an adjustable parameter in the theory since making $R$ large (small) provides a convenient way to study the qualitative effects of long (short)-range RKKY interaction on graphene magnetism.  Obviously, magnetism is strongly suppressed when $R$ is small. If experimental results on multilayer graphene magnetism become available in the future, it is straightforward to include quantitative effects of a $J$ and $T$ dependent mean-free path in our theory.

\begin{figure}
\includegraphics[width=1.0\linewidth]{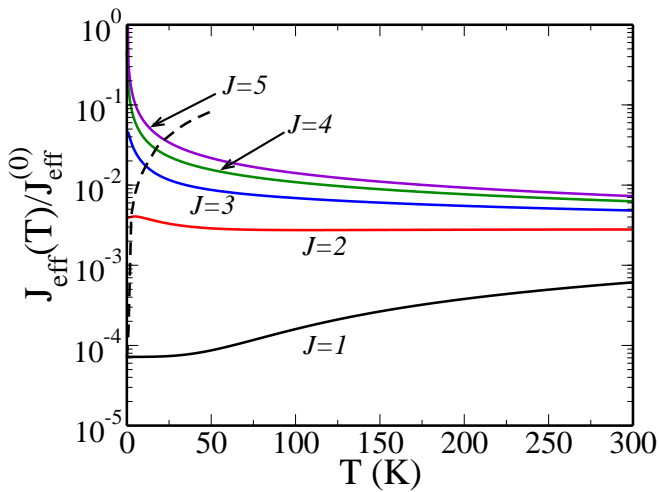}
\caption{The calculated effective coupling (solid lines) as a function of temperature for various chiralities $J=1,2,3,4,5$. In this calculation, $q_c^{(l)}=0.01/a$, $q_c^{(h)}=1/a$, and exponential cutoff $R=100 a$ are used.
Here the normalization factor $J_{\rm eff}^{(0)}={\left[J_{ex} a^2 \right]^2 } \times {2\pi D_1(a^{-1}) / 4\Omega_{\rm unit}}$ is independent of the chirality index $J$ and temperature $T$.  
The dashed line represents
$3k_BT/[S(S+1)x J^{(0)}_{\rm eff}]$ and the intersections with $J_{\rm eff}(T)$ indicate the transition temperatures solved self-consistently. 
} 
\label{fig:coupling_temperature}
\end{figure}

Figure \ref{fig:coupling_temperature} shows the calculated temperature dependence of the effective coupling for C2DESs with chiralities $J=1,2,3,4,5$. As shown in Fig.~\ref{fig:coupling_temperature}, the effective coupling for $J\ge 2$ decreases with temperature as in ordinary non-chiral 2DES,\cite{disorder1986} but 
for $J=1$ it increases with temperature, which is the direct consequence of the temperature dependence of the static susceptibility as shown in Fig.~\ref{fig:polarization}. The effective coupling also increases with increasing chiral index $J$ (or number of layers in rhombohedral multilayer graphene) because of the susceptibility behavior at long wavelength limit as shown in Fig.~1.
Obviously the $T$ and $J$ dependence of the effective magnetic coupling shown in Fig.~3 determines the magnetic transition temperature in DMG as discussed below.


From the calculated effective coupling we obtain the critical temperature for the magnetic transition in multilayer graphene. For the Heisenberg classical spins 
the mean-field transition temperature $T_{\rm c}$ is given by \cite{kittel_book,dassarmatc}
\begin{equation}
k_{\rm B}T_{\rm c}={S (S+1) \over 3} x J_{\rm eff},
\label{tc}
\end{equation}
where $S$ is the impurity spin, $x=n_{\rm imp} a^2$ is the concentration of the local moments with $n_{\rm imp}$ being the effective 2D magnetic impurity doping density. 
In the absence of any other information, we assume the magnetic dopants to be randomly distributed, but it is easy to include any correlations among the dopant positions if such dopant clustering effects are important.  We note that the ferromagnetic transition temperature is proportional to $J_{\rm ex}^2$ and $x$, but its dependence on the layer index $J$ is highly nontrivial and cannot be simply inferred using dimensional analysis since the layer index $J$ enters the DOS in a highly nonlinear manner [see Eq.~(4)].
We note that in our finite temperature RKKY model the ferromagnetic transition temperature  is obtained from Eq.~(\ref{tc}) by solving it self-consistently because $J_{\rm eff}$ itself is also strongly temperature dependent.\cite{dassarmatc} 
We emphasize that the strong temperature-dependence of the RKKY interaction in intrinsic graphene is the key physics determining the DMG ferromagnetic transition temperature in the theory.  If one makes the simplistic (and incorrect) assumption that $J_{\rm eff}$ is a temperature-independent coupling given by its $T=0$ value $J_{\rm eff}(0)$, then the transition temperatures are going to be unrealistically high.  The self-consistent solution of Eq.~(12) using the full temperature dependence of the magnetic coupling as shown in Fig.~3 is crucial in the theory to obtain the correct magnitude as well as the correct $J$-dependence of the transition temperature $T_c$.\cite{dassarmatc}

\begin{figure}[b]
\includegraphics[width=1.0\linewidth]{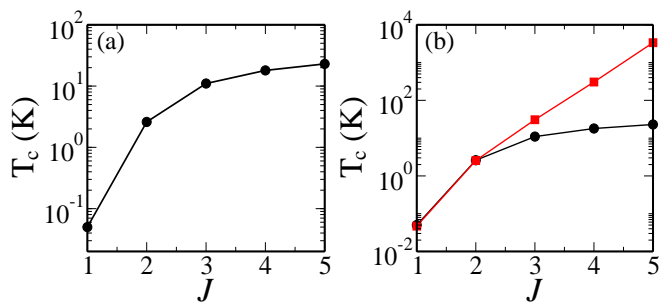}
\caption{The calculated ferromagnetic transition temperature as a function of layer (chiral) index $J$. We use the parameters $J_{\rm ex}=1$ eV, $S=5/2$, $x=0.05$, and $\Omega_{\rm unit}/a^2=1$ in this calculations. In (a) the self-consistent results are shown for different $J$. In (b), the self-consistent results (circles) are compared with the results (squares) obtained in the non-consistent method with $J_{\rm eff}(T=0)$.
} 
\label{fig:tc}
\end{figure}

In Fig.~\ref{fig:tc} we show the calculated self-consistent transition temperature $T_c$ as a function of layer index $J$. With the temperature dependent $J_{\rm eff}(T)$ and typical values of $J_{\rm ex}=1$ eV, $S=5/2$, $x=0.05$, $\Omega_{\rm unit}/a^2=1$, the self-consistent results show
the ferromagnetic transition temperatures $T_{\rm c}\approx 0.05, 2.6, 11, 18$ and $23$ K for $J=1,2,3,4,5$, respectively, [see Fig.~\ref{fig:tc}(a)]. In Fig.~\ref{fig:tc}(b) we compare them with the results calculated in the non-consistent method with $J_{\rm eff}(T=0)$ value
noting that the non-selfconsistent $T_c$ is unreasonably high.  Fig.~\ref{fig:tc}(b) shows that the zeroth order mean field results 
assuming $J_{\rm eff}$ to be given by its $T=0$ value
overestimate $T_c$ 
by an order of magnitude (or more)
for $J\ge3$ compared with the self-consistent results.
We note that $T_c$ is proportional to $xJ_{\rm ex}^2 S(S+1)$ and the results in Fig.~4 are for very specific values of $x$, $J_{\rm ex}$, and $S$, but one can scale the results to obtain $T_c$ for other values of $J_{\rm ex}$, $S$, and $x$.  We do emphasize, however, that $x$ cannot be too large so that one is in the dilute moment regime for the validity of our continuum theory.  For large impurity concentration ($x>0.1$ or so) DMG physics is different since graphene band structure itself may be affected.  In addition, the results obviously depend also on the basic graphene band parameters $g$, $v_0$, and $t_{\perp}$ [see Eq.~(4)] and this dependence is complex.  We choose the standard parameter values: $g=4$, $v_0=10^6$ m/s, and $t_{\perp}=0.3$ eV in our calculations.  We discuss the dependence on various cutoff parameters in the next section.
Our results apply to the rhombohedral stacking of graphene layers because the rhombohedral stacking sequence with $J$ layers is described by C2DES with the chirality index $J$. Thus, with high values of layer index $J$ the ferromagnetic ordering of magnetic impurities can be experimentally accessible in rhombohedral multilayer graphene
provided suitable magnetic dopants are used with reasonable ($\sim 1$ eV or so) local exchange coupling.

We note that the calculated $J$ dependence of the self-consistent $T_c$ in Fig.~4 is roughly linear whereas the corresponding dependence in the non-self-consistent mean field theory is nonlinear with a high power of $J$.  We do not believe that there is a generic unique power law behavior of $T_c$ on $J$, and the linear dependence in Fig.~4 applies only for our calculated results although it should be approximately valid for higher $J$ values.  Of course, $T_c$ is strongly suppressed for short mean free path due to disorder effects, which can only be discussed quantitatively for specific experimental situations.

\section{Discussion and conclusion}
We have studied the temperature dependence of the RKKY interaction and effective magnetic ordering as a function of layer number index for the C2DES of rhombohedrally stacked multilayer graphene. The chiral effective Hamiltonian used in this work is obtained from a perturbation theory taking into account only nearest-neighbor intralayer and interlayer hoppings\cite{min2008}, which is valid when we neglect the contributions from the trigonal warping terms which are much smaller than the terms kept in the effective Hamiltonian. 
Our theory is valid only when quantum fluctuations and direct exchange coupling between the impurity moments are negligible, which should be valid for large impurity spins and dilute impurity concentrations.
The valid energy scale for the chiral effective model is given by 0.03 eV--0.3 eV\cite{zhang2010}, which corresponds to the momentum scale of $0.01/a$--$0.1/a$. Thus it is natural to introduce the infrared low momentum cutoff and the ultraviolet high momentum cutoff, denoted by $q_c^{(l)}$ and $q_c^{(h)}$, respectively. 
These regularizations are necessary for obtaining meaningful results in graphene.

\begin{figure}[t]
\includegraphics[width=1.0\linewidth]{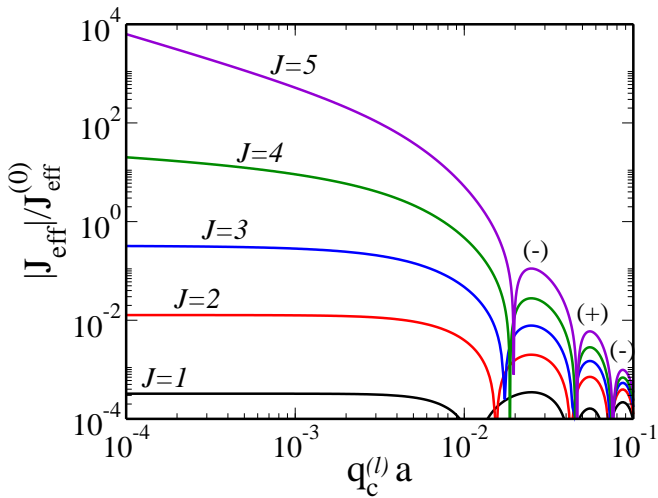}
\caption{The infrared momentum-cutoff dependence of the effective coupling for various chiralities $J=1,2,3,4,5$ with fixed high momentum cutoff $q_c^{(h)}=1/a$ and exponential cutoff $R=100 a$.} 
\label{fig:cutoff_low}
\end{figure}

In our model, the calculated effective coupling $J_{\rm eff}$ is insensitive to the ultraviolet cutoff $q_c^{(h)}$ for $J \ge 2$, while for $J=1$, $J_{\rm eff}$ shows the well-known logarithmic ultraviolet divergence at high momenta. Note that for monolayer graphene ($J=1$ C2DES) there is no interlayer hopping and the valid momentum scale is restricted only by the inverse lattice constant, beyond which the linear dispersion is no longer valid. 
Even though our results are independent of the high momentum cutoff, they are affected by the low momentum cutoff $q_c^{(l)}$. As shown in Fig.~\ref{fig:cutoff_low}, the calculated effective couplings are consistent for small values of the cutoff $q_{c}^{(l)}< 10^{-2}/a$. However, for large values of cutoff the sign of $J_{\rm eff}$ oscillates with $q_c^{(l)}$. For a typical value of $q_c^{(l)}=0.01/a$, $J_{\rm eff}>0$, and thus the ordering is ferromagnetic for all C2DESs. 
Note that the calculated results also depend on the exponential disorder cutoff $R$ for $R> 500a$, and a larger $R$ gives more oscillating behavior in $J_{\rm eff}$. 
The finite mobility of multilayer graphene, however, restricts the size of $R$ and for a typical scale of mean-free path the results do not change qualitatively. In addition, the finite mean free path cutoff prevents the system from becoming an interaction-induced ordered state with a non-zero energy gap at the Dirac point, thus we can use a chiral gas model of a gapless semimetal even at zero carrier density.

In summary, we study the magnetic properties of multilayer graphene (chiral 2D electron systems) in the presence of magnetic impurities
as a function of layer index number in the intrinsic semimetallic situation.
By calculating the temperature dependent susceptibility of multilayer graphene we investigate the temperature dependence of the RKKY interaction and the associated carrier induced effective magnetic coupling using the effective chiral model of multilayer graphene. We show that due to the enhanced DOS in rhombohedral stacking the ferromagnetic ordering between magnetic impurities is possible at experimentally accessible temperatures. Our results indicate that
the magnetic impurity induced ferromagnetic order in multilayer graphene
should be observable experimentally for layer number 3 or above in multilayer graphene system, perhaps ushering in the physics of spintronics based on diluted magnetic graphene.

\section*{ACKNOWLEDGMENTS}
This work is supported by LPS-MPO-CMTC.


\end{document}